\documentclass[prb]{revtex4}%
\usepackage{amsfonts}
\usepackage{amsmath}
\usepackage{amssymb}
\usepackage{graphicx}%
\begin{document}
\title{Carrier relaxation dynamics in high temperature superconductors}
\author{Jure Demsar$^{\ast}$}
\affiliation{Los Alamos National Laboratory, MS K764, Los Alamos, NM 87545, USA}
\author{Dragan Mihailovic, and Viktor V. Kabanov  }
\affiliation{"J. Stefan" Institute, Jamova 39, SI-1000, Ljubljana, Slovenia}

\begin{abstract}
In this report we review recent experimental results on photoexcited carrier
relaxation dynamics on high temperature superconductors (HTSC) probed by a
femtosecond time-resolved optical spectroscopy, and compare the results with
the data obtained on quasi two dimensional charge density waves. In these
experiments, a femtosecond laser pump pulse excites electron-hole pairs via an
inter-band transition in the material. These hot carriers rapidly release
their energy via electron-electron and electron-phonon collisions reaching
states near the Fermi energy within $\sim$100 fs. If an energy gap is present
in the low-energy density of states (DOS), it inhibits the final relaxation
step and photoexcited carriers accumulate above the gap causing a transient
change in reflectivity arising from excited state absorption. The relaxation
and recombination processes of photoexcited quasiparticles, governed by the
magnitude, anisotropy and the T-dependence of the low energy gap, are
monitored by measuring the resulting photoinduced absorption as a function of
time after the photoexcitation. This way, the studies of carrier relaxation
dynamics give us direct information of the T-dependent changes in the low
energy DOS. The technique is particularly useful to probe the systems with
spatial inhomogeneities, where different local environments give rise to
different relaxation rates. The data on series of HTSC-s show evidence for the
coexistence of two distinct relaxation processes, whose T-dependences seem to
be governed by two different energy scales: a T-independent pseudogap and a
mean-field-like T-dependent gap that opens at T$_{c}$. The data suggest the
origin of the two-gap behavior is in the intrinsic microscopic spatial
inhomogeneity of these materials.

\end{abstract}
\maketitle

\section{INTRODUCTION}

Femtosecond time-resolved optical spectroscopy has been shown in the last
couple of years to present an excellent alternative to more conventional
time-averaging frequency-domain spectroscopies for probing the changes in the
low energy electronic structure in strongly correlated systems
\cite{001Kabanov,002ACS}. In these experiments, a femtosecond laser pump pulse
excites electron-hole pairs via an interband transition in the material. In a
process which is similar in most materials including metals, semiconductors
and superconductors, these hot carriers rapidly release their energy via
electron-electron and electron-phonon collisions reaching states near the
Fermi energy within 10-100 fs. Further relaxation and recombination dynamics,
determined by measuring photoinduced changes in reflectivity or transmission
as a function of time after photoexcitation, depends strongly on the nature of
the low-lying electronic spectrum. In particular, the experimental technique
was found to be sensitive to opening of the superconducting gap,$^{1-18}$
appearance of a short-range and long range charge-density wave order
\cite{005TRbb,006TR2DCDW}, and changes in the electronic specific heat
associated with the heavy fermion behavior \cite{021HF}, just to mention a
few. What is particularly important is the fact, that even though the probe
photon wavelength in these experiments ranges from THz
\cite{008Averitt,022JDMgB2} (enabling direct measurement of photoinduced
conductivity dynamics),
mid-IR\cite{001Kabanov,0099Bi2212,010Tl2201,013Hg1223,007Segre} to several
eV\cite{015Stevens}, the relaxation dynamics is the
same\cite{008Averitt,022JDMgB2}, giving further support to the basic idea
\cite{001Kabanov} that the photoinduced reflectivity (transmissivity) dynamics
is governed by relaxation and recombination processes of quasiparticles in the
vicinity of Fermi energy.

Since the optical penetration depth in these materials is on the order of 100
nm, the technique is essentially a bulk probe. Moreover, since the effective
shutter speed is on the order of a picosecond, the technique is particularly
useful to probe the systems with (dynamic) spatial inhomogeneities. In this
case different local environments (that appear frozen on the timescale of
picoseconds) give rise to different components in measured photoinduced
reflectivity (transmissivity) traces. As the different components can have
different time scales\cite{014Y124,019ODpaper},
temperature\cite{001Kabanov,005TRbb}, photoexcitation intensity, and probe
polarization\cite{014Y124} or wavelength dependences, they can be easily extracted.

In this report we summarize some of the recent results on various
high-temperature superconductors (HTSC), where the data show evidence for the
coexistence of two distinct relaxation processes whose T-dependences seem to
be governed by different energy gaps: a T-independent pseudogap and a
mean-field-like T-dependent gap that opens at T$_{c}$. The data suggest the
origin of the two-gap behavior is in the intrinsic microscopic spatial
inhomogeneity of these materials. In additions we present experimental data on
two prototype quasi two dimensional (2D) charge density wave (CDW)
dichalcogenides 1$T$-TaS$_{2}$ and 2$H$-TaSe$_{2}$, and compare these results
with the data on HTSC.

\section{EXPERIMENTAL DETAILS}

In these experiments, a Ti:sapphire mode-locked laser operating at a 78
MHz\ repetition rate and pulse length of 50 fs was used as a source of both
pump and probe pulse trains. The wavelength of the pulses was centered at
approximately $\lambda$ $\approx$ 800nm (1.58eV) and the intensity ratio of
pump and probe pulses was approximately 100:1. The pump and probe beams were
crossed on the sample's surface, where the angle of incidence of both beams
was less than 10$^{o}$. The diameter\ of \ both beams on the surface was
$\sim$100$\mu$m. The typical energy density of pump pulses was $\sim1$ $\mu
$J$/$cm$^{2}$, which produces a weak perturbation of the electronic system
with approximately 3$\times10^{10}$ thermalized photoexcited carriers per
pulse (the approximation is based on the assumption that each photon creates
$\hbar\omega/\Delta$ thermalized photoexcited carriers, where $\Delta
\approx40$ meV is of the order of the superconducting or CDW gap
\cite{001Kabanov}). The train of the pump pulses was modulated at 200kHz
\ with an acousto-optic modulator and the small photoinduced changes in
reflectivity or transmission were resolved out of noise with the aid of
phase-sensitive detection. The pump and probe beams were cross-polarized to
reduce scattering of pump beam into the detector (avalanche photodiode).\ A
detailed description of the experimental technique can be found in Ref. [2].

\section{CARRIER\ DYNAMICS IN\ CUPRATES AND\ 2-D CHARGE\ DENSITY\ WAVES}

In this section we discuss some recent experimental results on carrier
relaxation dynamics on HTSC and low dimensional CDW's with particular focus on
comparison of the dynamics in the two systems and its implications. The
comparison of carrier relaxation dynamics between the two systems is
particularly interesting, since HTSC and quasi-2D CDW's are thought to exhibit
some important similarities. Both are layered, highly anisotropic materials
which are often described in terms of a quasi-2D Fermi surface (FS) in their
normal state. In HTSCs, it is commonly believed that the superconducting gap
has nodes along certain directions on the FS due to the $d$-wave component of
the order parameter, whereas in 2D-CDW systems a CDW\ gap is also expected
only along certain wavevectors, remaining gapless (and metallic) on other
regions of the FS. The low-energy single particle excitations in the two
classes of compounds might therefore be expected to show some important common
features related to reduced dimensionality. However, the validity of the
Fermi-liquid (FL) concept when applied to low-temperature properties in HTSCs
has repeatedly been brought into question, suggesting that new insight into
the physics of quasi-2D systems may be gained by investigating the low-energy
electronic gap structure and carrier recombination dynamics of quasi-2D CDW
dichalcogenides with femtosecond spectroscopy.

Figure 1 shows the raw photoinduced reflectivity data on high temperature
superconductor YBa$_{2}$Cu$_{3}$O$_{7-\delta}$ (YBCO), compared to two quasi
two dimensional charge density wave systems 1T-TaS$_{2}$ and 2H-TaSe$_{2}$.
The data are plotted with solid symbols, together with solid lines
representing the best fit obtained using the function $\Delta R/R=f(t)[A\exp
(-t/\tau_{1})+B\exp(-t/\tau_{2})+..+D]$. Here $f(t)=H(t)[1-\exp(-t/\tau
_{ee})]$ represents the finite risetime, with $H(t)$ being the convoluted
Heaviside step function accounting for the finite laser pulse width
\cite{002ACS}, and $\tau_{ee}$ is the electron-electron thermalization time.
The decay dynamics is described by one or more exponential decays with
corresponding amplitudes $A$, $B$,.. and relaxation times $\tau_{1}$,
$\tau_{2}$..\ and a possible long lived component with amplitude $D$ (lifetime
of the long lived component is shown to be longer then the separation between
two successive excitation pulses of 12 ns \cite{016KabanovSlow}). In the case
of charge density waves, the oscillatory component with frequency $\omega$ and
damping constant $\Gamma$ has been observed, superimposed on the decaying
transient \cite{005TRbb,006TR2DCDW} - see Fig. 1 b) and c). The oscillatory
transient was found to be due to coherently excited\cite{Zeiger} collective
mode - amplitudon, in very good agreement with Raman spectroscopy data
\cite{005TRbb,006TR2DCDW}, and will not be discussed in detail here. In this
report we focus on the picosecond decay dynamics, described by the
temperature, photoexcitation intensity and the probe polarization dependence
of amplitudes ($A$,$B$..) and relaxation times ($\tau_{1}$, $\tau_{2}$..),
whose temperature dependences are shown in Fig. 2.

Figure 2 a) and d) presents the generic temperature dependence of relaxation
times and amplitudes of the photoinduced transient obtained on HTSC over the
last decade or so \cite{001Kabanov,003Han,004Eesley,013Hg1223,019ODpaper}. At
temperatures below T$_{c}$, the relaxation dynamics is showing a two-component
behavior, with the two components having the
same\cite{019ODpaper,013Hg1223,014Y124} or opposite sign
\cite{004Eesley,010Tl2201,014Y124}, depending on the probe photon wavelength
and/or polarization of the probe light with respect to the crystal
axis\cite{014Y124}. The component that is present only below T$_{c}$ shows
vanishing amplitude (dashed line in Fig. 2 d)) and diverging relaxation time
(Fig. 2 a)) as the temperature approaches T$_{c}$. This behavior is attributed
to opening of the superconducting gap and corresponding quasiparticle (QP)
relaxation bottleneck\cite{001Kabanov} that occurs due to the fact that there
are no ingap states available in the process of QP recombination to the
condensate. The model\cite{001Kabanov} is based on the fact that near T$_{c}$
the gap is small and the number of thermally excited QPs is large compared to
the number of photoexcited ones. In this case the process of recombination of
two QPs to the condensate is fast. However, since recombination of two QPs
creates high energy phonon, which can further excite QP pairs, the relaxation
dynamics near T$_{c}$ is governed by anharmonic phonon decay. As the gap
$\Delta(T)$ is closing, the specific heat of phonons with energy less than the
gap energy is becoming smaller, and the relaxation time increases as
$\tau\propto1/\Delta(T)$ \cite{001Kabanov}. The relaxation time of the second
component has been found to show only weak temperature dependence. The studies
of T-dependence of the amplitude of the transient as a function of doping have
revealed, that it is associated with the pseudogap energy scale\cite{024EPL}.
The coexistence of the two relaxation component below T$_{c}$ implies some
kind of microscopic phase separation, with coexisting areas of high carrier
and low carrier densities\cite{019ODpaper}.

In order to test the predictions of the theoretical model\cite{001Kabanov} and
the implications of the experimental results on the physics of high
temperature superconductors\cite{001Kabanov,019ODpaper}, we have applied the
same technique to study 1D-CDW \cite{005TRbb} and quasi 2D-CDW
compounds\cite{006TR2DCDW}. In the following we discuss the observed
T-dependences of the relaxation dynamics in 1$T$-TaS$_{2}$ and 2$H$-TaSe$_{2}$
in view with the spectroscopic data on associated changes in the low energy
density of states.

At room temperature 1$T$-TaS$_{2}$ is in a nearly-commensurate (\textit{nc}-)
CDW phase. Around $T_{nc-c}=200$ K it undergoes a strongly first-order
\textquotedblright lock-in\textquotedblright\ transition to a \textit{c}-CDW
state\cite{Review}. In spite of the expected appearance of a gap in parts of
the Fermi surface due to nesting at $T_{nc-c}$, the whole FS was found to
exhibit a \textquotedblright pseudogap\textquotedblright\ feature already at
room temperature with a finite density of states (DOS) at $E_{F}$
\cite{DardelPilloManske}. Upon lowering the temperature through $T_{nc-c}$, a
further \emph{abrupt} decrease in the DOS is observed near $E_{F}$,
accompanied by an order of magnitude \textit{increase }in resistivity. Yet in
spite of the presence of a CDW\ gap at low temperatures 1$T$-TaS$_{2}$ is
reported to have a small but finite DOS at $E_{F}$ in the low-temperature
\textit{c}-phase \cite{DardelPilloManske}. The picosecond relaxation dynamics
in 1$T$-TaS$_{2}$ requires a \emph{stretch }exponential decay fit, $A$
$exp(-(t/\tau)^{p})$ , with $p\sim0.5$ to fit the data adequately. (Although
the SP transient in 1$T$-TaSe$_{2}$ can also be fit by the sum of two
exponentials, the two components have the same $T$-dependence, which suggests
that we are dealing with a single relaxation process with non-exponential
dynamics rather than two distinctly independent processes). The observation of
a stretch exponential decay - which typically describes systems with a spread
of relaxation times - is consistent with the observed finite DOS at E$_{F}$
below $T_{c-nc}$ \cite{DardelPilloManske}. Since $\tau\sim1/\Delta$ [1] the
observed stretch exponential decay actually implies a near-Gaussian spread of
$1/\Delta$. The $T$-dependences of the amplitude and $\tau$ (using $s=0.5$)
for 1$T$-TaS$_{2}$ are plotted in Fig. 2 e) and b), respectively. The
relaxation dynamics are clearly strongly affected by the lock-in transition
around 200K. We observe an abrupt hysteretic change of both amplitude and
$\tau$ around $T_{nc-c},$ consistent with an abrupt appearance of a gap at
$T_{nc-c}$ suggested by other experiments \cite{DardelPilloManske}.\ Upon
further cooling amplitude remains more or less constant, while $\tau$ slowly
decreases. Upon warming, a rapid drop in amplitude and $\tau$ associated with
gap closure now occurs at around 220 K, consistent with the hysteresis
observed in transport measurements\cite{Review}. Above~230 K the photoinduced
transient is fast and very weak.

2$H$-TaSe$_{2}$ on the other hand is expected to bear close resemblance to
cuprates. It exhibits metallic properties above room temperature. Upon cooling
it undergoes a second order phase transition to an incommensurate
(\textit{i}-) CDW state at $T_{n-i}=122$ K. This phase transition is
reportedly accompanied by the appearance of a gap on the Fermi surface (FS)
centered at the K point, but apparently - according to photoemission
studies\cite{Liu} - remains gapless on the part of the FS centered at the
$\Gamma$ point. The transition is accompanied by a decrease in the scattering
rate and a corresponding \emph{drop} in resistivity \cite{Vescoli}. The onset
of a \textit{c}-CDW phase at $T_{i-c}=88K$ leaves the excitation spectrum as
well as the transport and thermodynamic properties almost unaffected
\cite{Review,Liu}. Since below $T_{n-i}$, the system is expected to have a
highly anisotropic gap with gapless regions over parts the FS\cite{Liu}, the
relaxation dynamics in 2$H$-TaSe$_{2}$ is expected to be non-exponential
as\ in 1$T$-TaS$_{2}$. However, the relaxation dynamics in 2$H$-TaSe$_{2}$
over the entire $T$-range can be described well by the single-exponential
decay. Moreover, below $T_{n-i}$ both the $T$-dependence of amplitude and
relaxation time $\tau$ obtained from the single exponential fit to the data on
2$H$-TaSe$_{2}$, plotted in Fig. 2 f) and c), show extremely good agreement
with the prediction of the model for carrier relaxation across a well-defined
temperature-dependent gap \cite{001Kabanov}. The theoretical fit to the data
using the model by Kabanov et al. \cite{001Kabanov} with a BCS T-dependence of
the gap is superimposed (lines), giving the value of the gap of $2\Delta
(0)=70\pm10$ meV. The value is somewhat smaller than the maximum gap obtained
from tunneling \cite{tunelling}, but in good agreement with the maximum gap
value of 65 meV from ARPES\cite{Liu}. Above $\sim140$ K the relaxation time
and $S(T)$\ show only a weak $T$-dependence with $\tau\sim0.1-0.17$ ps. This
time is close to the expected electron-phonon thermalization time (in case
there is no gap in the DOS, like in metals, the photoinduced
reflectivity/transmissivity dynamics is governed by energy transfer from
thermalized electron subsystem to the lattice).\cite{Ippen,Allen} On the other
hand, the behavior of amplitude\ just above $T_{i-n}=120$ K is rather unusual,
showing a rapid increase with increasing temperature between $120$ and
$\sim140$ K. This could be attributed to the presence of short-range segments
of ordered CDW, as observed in quasi-1D CDW K$_{0.3}$MoO$_{3}$\cite{005TRbb}.

The emerging picture based on the time-domain measurements on HTSC and
2H-TaSe$_{2}$ presented here is one in which the low-temperature state shows
\emph{a clear large gap }in the excitation spectrum on the femtosecond
timescales (not just a depression in the DOS such as is observed in
time-averaged experiments). There is also clear evidence for very slow
relaxation of excitations whose energy is less than the maximum gap - in both
HTSC\cite{016KabanovSlow,018Feenstra} and 2H-TaSe$_{2}$.\cite{006TR2DCDW} The
observed behavior is in clear contradiction with a FL interpretation, where
the QP\ relaxation would be expected to occur primarily in the gapless regions
of the FS (in the nodes for the case of superconductors). The observation of
only a large gap on the femtosecond timescale implies that there are certain
momenta associated with the gapless regions which are either inaccessible to
quasiparticles, or - implying a breakdown of the FL picture altogether -
simply that \emph{extended} states with these wavevectors do not
exist\emph{\ }at all. The latter behavior is consistent with the notion of
fluctuating \emph{locally} \emph{ordered }regions in space, in which case it
becomes clear why one cannot speak of FL-like quasiparticle excitations with
well defined momenta. The precursor \textquotedblright pseudogap
state\textquotedblright\ appears to be associated with the fluctuating
presence of fully gapped short-range-ordered CDW patches or segments, similar
to the locally gapped regions in real space arising from a statistically
fluctuating population of pre-formed pairs in
HTSCs\cite{001Kabanov,019ODpaper}.


\section{LOW TEMPERATURE RELAXATION TIME IN CUPRATES}

In this section we analyze the relaxation dynamics in high temperature
superconductors at low temperatures (T$\ll$T$_{C}$) and low excitation
fluences, that has gained considerable interest in the last couple of years
\cite{007Segre,013Hg1223,0099Bi2212,010Tl2201,023Schneider}. It has been shown
that at extremely low fluences the low temperature relaxation time depends on
photoexcitation intensity ($\emph{P}$)\cite{007Segre}, and in the limit of
$\emph{P}\rightarrow0$ the relaxation time diverges as $T\rightarrow0$ K. This
has been consistently observed in the extremely low excitation regime on
Bi2212\cite{0099Bi2212}, Tl2201\cite{010Tl2201}, Hg1223\cite{013Hg1223},
underdoped YBCO\cite{007Segre}, Bi2201\cite{023Schneider} and
LASCO\cite{023Schneider}.

We interpret the low temperature (low photoexcitation fluence) increase of
relaxation time in terms of QP \textit{recombination} bottleneck. Namely, the
assumption that the two-particle recombination is fast compared to the
anharmonic phonon decay (that determines the decay of photoexcited population
density ($n_{pe}$) at temperatures close to T$_{c}$), can be violated at low
temperatures, when the gap is large and the density of thermally excited QPs
($n_{th}$) is small. It can lead to a situation when the recombination time
becomes longer than the anharmonic phonon decay time, and in this case the
relaxation time of photoexcited QP density is governed by bi-particle
\textit{recombination} process. To show this, we have to bare in mind that in
this low excitation fluence experiments the number of thermally excited QPs
is,\ except at very low temperatures, much larger than the number of
photoexcited ones. In this limit, the rate equation describing the dynamics of
QP density $n$ can be written as%

\begin{equation}
\frac{\partial n}{\partial t}=-\beta\left(  n^{2}-n_{th}^{2}\right)  \text{
\ ; \ }n=n_{th}+n_{pe}\text{ \ ,} \label{Eq1}%
\end{equation}
where $\beta$ is a constant. Due to the fact, that at low excitation densities
$n_{th}\gg n_{pe}$ in most of the temperature range, we can linearize the
equation to obtain
\begin{equation}
\frac{\partial n_{pe}}{\partial t}=-\frac{n_{pe}}{\tau_{R}}\text{ \ \ ;
\ \ }\tau_{R}=\left(  2\beta n_{th}\right)  ^{-1}\text{ .} \label{Eq2}%
\end{equation}
Since $n_{th}\propto\exp(-\Delta/T)$, where $\Delta$ is the superconducting
gap, the relaxation time is expected to increase exponentially at low
temperatures. In the limit where $n_{th}\gg n_{pe}$ the relaxation dynamics is
expected to be exponential. On the other hand, at very low temperatures, when
$n_{pe}$ $\gg n_{th}$, the Eq.(\ref{Eq1}) cannot be linearized and the
dynamics is non-exponential. More rigorously, the relaxation time in the limit
$n_{th}\gg n_{pe}$ and $T\ll T_{c}$ can be obtained explicitly by considering
the kinetic equation for QPs \cite{013Hg1223,Aronov}:
\begin{equation}
\tau_{R}=\frac{\hbar\Omega_{c}^{2}}{32\pi\lambda\Delta^{2}\sqrt{\pi\Delta
k_{B}{T}/2}}\exp{(\Delta/k}_{B}{T)\;.} \label{Vfin}%
\end{equation}
Here $\lambda$ is the dimensionless electron-phonon coupling constant, for
HTSC typically on the order of 1 \cite{Brorson}, and $\Omega_{c}$ is a typical
phonon cutoff frequency. There is a subtle point to Eq.(\ref{Vfin}) that needs
to be mentioned: temperature ${T}$ in Eq.(\ref{Vfin}) is actually the
temperature of phonons with energy less than $2\Delta$, and can be slightly
higher then the equilibrium sample temperature. These corrections can be
important at very low temperatures (see Ref. [9] for details).

Fig. 3 a) shows the raw photoinduced reflectivity data (symbols) on Hg1223 at
different temperatures, together with fits (solid lines) using a single
exponential decay. We should note that even though the lowest sample holder
temperature in the experiments was 4 K, the actual lowest temperature in the
illuminated spot is substantially higher ($\sim40$ K) due to heating induced
by excitation pulse train. The temperature increase is determined by thermal
conductivity of the sample, and can be easily accounted for \cite{002ACS},
giving the uncertainty in temperature of $\pm2$ K. The extracted relaxation
time as a function of temperature is plotted in Fig. 3 b) using solid symbols,
compared to the theoretical predictions for $\tau$ governed by recombination
bottleneck [Eq.(\ref{Vfin})] - solid line, and $\tau$ determined by phonon
bottleneck (Eq.(28) of Ref.[1]) - dashed line. As we can see, the
recombination bottleneck describes well the low temperature behavior. At
higher temperatures a crossover to phonon bottleneck determined relaxation is
expected. A crossover from high temperature \textit{relaxation} to low
temperature \textit{recombination} picture is expected to highly depend on the
magnitude of the superconducting gap $\Delta$. Since the gap value in YBCO,
determined from tunneling data is lower than that of Bi2212 (and Hg1223) the
crossover in YBCO is expected to be lower in temperature (or photoexcitation
intensity). At lower temperatures, when $n_{pe}\gtrsim n_{th}$, the relaxation
should be non-exponential. Indeed the crossover to non-exponential relaxation
was reported at very low temperatures in Bi2212 \cite{0099Bi2212} and Tl2201
\cite{011Smith}.

We should mention an alternative model for carrier relaxation dynamics in
cuprates recently suggested by Segre et al.\cite{007Segre}. Analyzing the
dependence of $\tau$ on temperature and photoexcitation intensity \emph{P},
the authors argue that the decay in $\Delta R/R$ is determined by QP
thermalization, rather than recombination \cite{001Kabanov}. This is clearly
inconsistent with published data on ultrafast conductivity dynamics on YBCO in
the THz spectral region\cite{008Averitt}, where the recovery of the condensate
due to recombination of QPs has been measured directly, and was found to be on
a picosecond time-scale. In fact, the T-dependence of the condensate recovery
time on optimally doped YBCO was found to be identical to the recovery time
extracted from all-optical pump-probe experiments at 1.5
eV\cite{001Kabanov,003Han,008Averitt}. Moreover, similar one-to-one relation
between the relaxation dynamics measured with the two experimental techniques
has recently been found also on MgB$_{2}$.\cite{022JDMgB2} Furthermore, in
Fig.3 of Ref.[16] it is shown that the relaxation rate ($1/\tau$) is
proportional to the amplitude of the signal. Such behavior is well known in
superconductors and arises from QP recombination to the condensate. Originally
it was discussed on a phenomenological level by Rothwarf and Taylor\cite{roth}
and later derived from kinetic equations for
superconductors\cite{013Hg1223,ovchkres}. We believe that the linear intensity
dependence of the relaxation rate is consistent with a QP recombination
mechanism\cite{001Kabanov} presented above, rather than QP thermalization, as
proposed by Segre et al.\cite{007Segre}.


\section{CONCLUSIONS}

The time-resolved experiments on high temperature superconductors and quasi 2D
charge density waves show that irrespective of the fundamental underlying
cause for the instability, these quasi-2D materials, show a transition from a
high-temperature uniform metallic state to a low-temperature correlated state
via the formation of a dynamically inhomogeneous intermediate state. Since the
timescale of the measurement is on the order of a picosecond, and the
inhomogeneities appear to be frozen on this timescale, different local
environments give rise to different relaxation dynamics. In this way these
measurements on high temperature superconductors reveal the coexistence of
high and low carrier density regions, with the recovery dynamics being
governed by superconducting gap and pseudo-gap respectively. Since the lateral
dimension of the inhomogeneities, as determined for example by EXAFS, is on
the order of several unit cells, the interpretation of the data in terms of
quasi-2D Fermi liquid is questionable. In fact, the observation of a clear
large gap in these experiments implies that there are certain momenta
associated with the gapless regions which are either inaccessible to
quasiparticles, or simply that \emph{extended} states with these wavevectors
do not exist\emph{\ }at all. The time-averaged response (such as is observed
in ARPES or infrared spectra) may then be thought of as the superposition of
the different components in the inhomogeneous state, while the observed
anisotropy reveals the directionality of the interaction which leads to the
formation of long range order.

Carrier dynamics studies on quasi-2D charge density waves have been performed,
to elucidate the carrier relaxation dynamics in cuprates. We confirm, that the
dynamics is sensitive to the opening of the gap in DOS. In 1T-TaS$_{2}$ the
relaxation dynamics is non-exponential, consistent with the spectroscopic data
showing finite DOS at E$_{F}$ at all temperatures. On the other hand, single
exponential decay dynamics, and the T-dependence of amplitude and relaxation
time in 2H-TaSe$_{2}$ suggests similar behavior as in cuprates, where low
temperature correlated state is formed via local precursor CDW segments.

Finally, we analyze the low-temperature, low excitation fluence dynamics in
cuprates, where the relaxation time was found to increase exponentially as
$T\longrightarrow0$. We show, that in this limit the recovery dynamics of
photoexcited quasiparticle density is governed by bi-particle recombination process.

\acknowledgments

We wish to acknowledge our collaborators Janusz Karpinski, Laszlo Forro,
Helmut Berger, Thomas Wolf, Airton A. Martin, Katica Biljakovic and Jan Evetts
for supplying the samples used in these investigations and for all the
valuable comments and discussions.


\section{\bigskip Figure Captions}

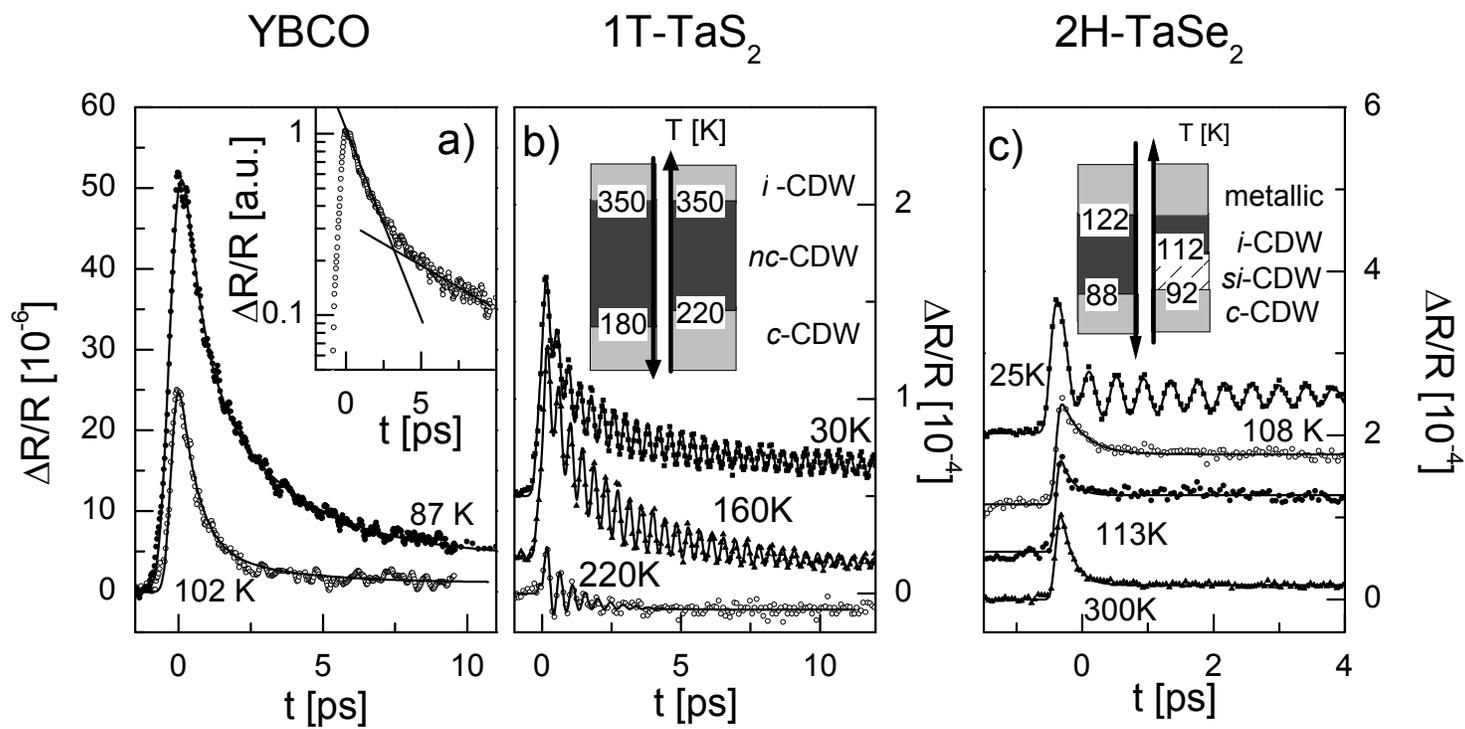
\textbf{Figure 1}: Photoinduced reflectivity traces taken at different
temperatures for a) optimally doped YBCO, b) 1T-TaS$_{2}$, and 2H-TaSe$_{2}$.
In b) and c) the traces are offset for clarity. Inset to panel a): the
two-exponential decay dynamics revealed on a semi-log scale. Insets to panels
b) and c): The phase diagrams of\ bulk 1$T$-TaS$_{2}$\ and 2$H$-TaSe$_{2}$.

\bigskip

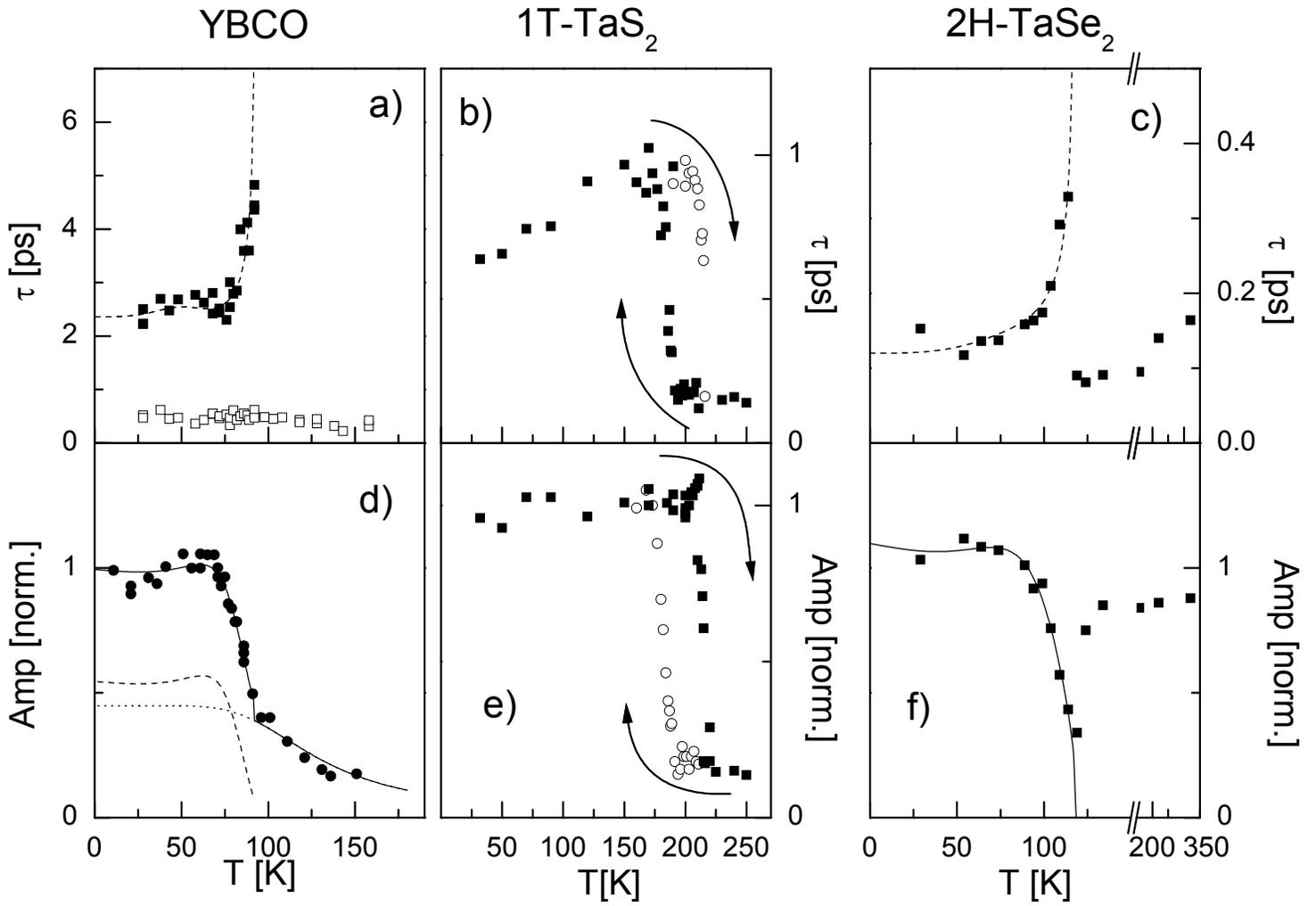
\textbf{Figure 2}: Panels a) $-$ c): the temperature dependences of picosecond
relaxation times of optimally doped YBCO, 1T-TaS$_{2}$, and 2H-TaSe$_{2}$.
Panels d) $-$ f): the corresponding T-dependences of amplitudes of
photoinduced changes. Lines are fits to the data using the model
\cite{001Kabanov} - see text. Note the hysteresis observed in T-dependence of
photoinduced transient on 1T-TaS$_{2}$.

\bigskip

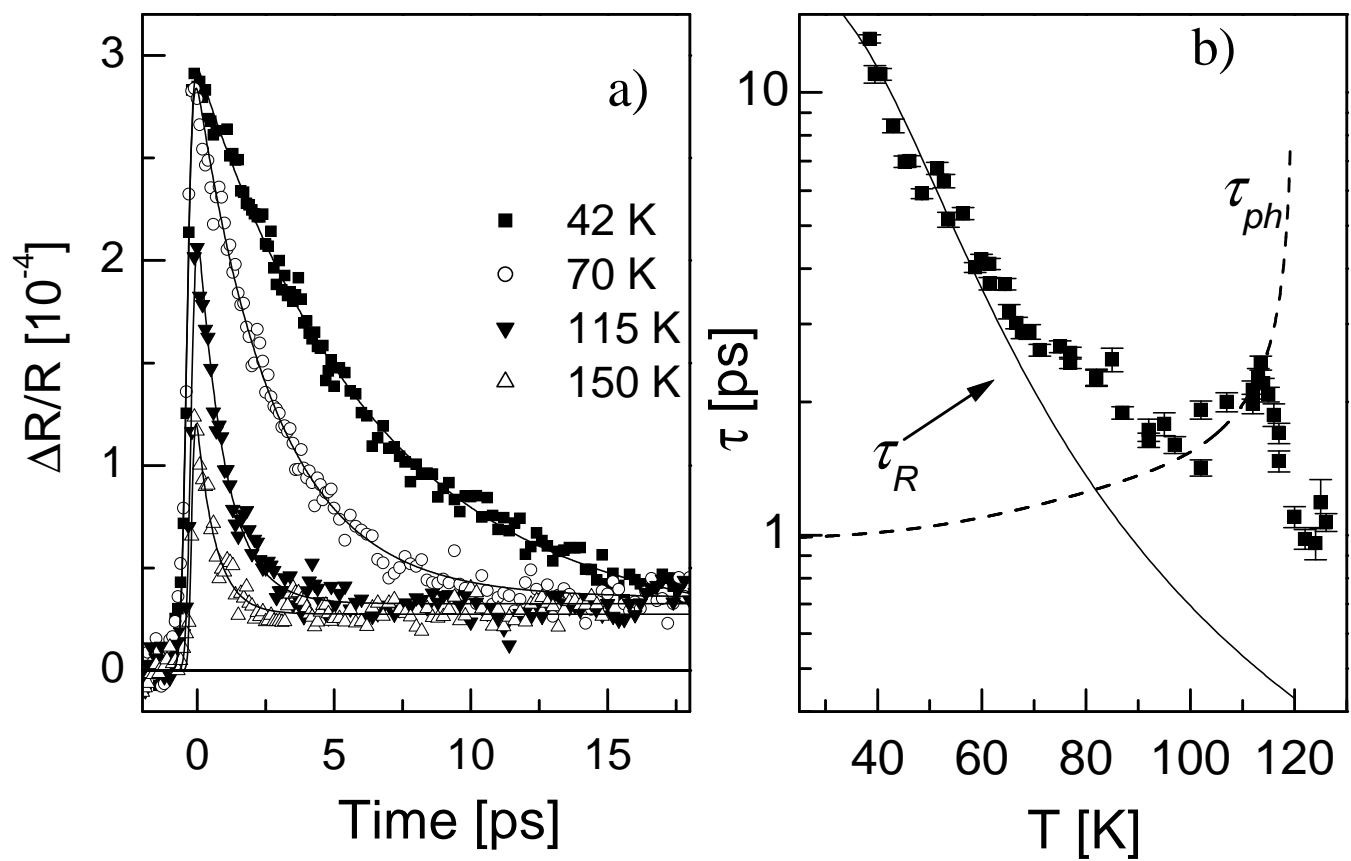
\textbf{Figure 3}: a) The photoinduced reflectivity on Hg1223 taken at various
temperatures elow and above T$_{c}$ = 120 K. Solid lines are fits to the data
using single exponential decay. b) The temperature dependence of the
relaxation time $\tau$ taken on Hg1223, compared to the theoretical curves for
bi-particle recombination bottleneck model (solid curve) and phonon bottleneck
model \cite{001Kabanov} - dashed curve.

\bigskip

* jdemsar@lanl.gov; phone 1 505 665-8839; fax 1 505 665-7652; Los Alamos
National Laboratory, Mail Stop K764, Los Alamos, NM, USA 87545

\end{document}